\documentclass[prb,twocolumn,superscriptaddress,showpacs,preprintnumbers,amsmath,amssymb]{revtex4}

\usepackage{graphicx}
\usepackage{dcolumn}
\usepackage{bm}

\begin{document}

\bibliographystyle{apsrev} 

\title{Protection of center-spin coherence by a dynamically polarized nuclear spin
core}

\author{Wenxian Zhang}
\affiliation{Key Laboratory of Micro and Nano Photonic Structures (Ministry of Education), Department of Optical Science and Engineering,
Fudan University, Shanghai 200433, China}

\author{Jian-Liang Hu}
\affiliation{Department of Physics, The Chinese University of Hong Kong,
Shatin, N. T., Hong Kong, China}

\author{Jun Zhuang}
\affiliation{Key Laboratory of Micro and Nano Photonic Structures (Ministry of Education), Department of Optical Science and Engineering,
Fudan University, Shanghai 200433, China}

\author{J. Q. You}
\affiliation{Department of Physics, Fudan University, Shanghai 200433, China}

\author{Ren-Bao Liu}
\affiliation{Department of Physics, The Chinese
University of Hong Kong, Shatin, N. T., Hong Kong, China}

\date{\today}

\begin{abstract}
Understanding fully the dynamics of coupled electron-nuclear spin systems, which are important for the development of long-lived qubits based on solid-state systems, remains a challenge. We show that in a singly charged semiconductor quantum dot with inhomogeneous
hyperfine coupling, the nuclear spins relatively strongly coupled to the
electron spin form a polarized core during the dynamical polarization
process. The polarized core provides a protection effect against the electron
spin relaxation, reducing the decay rate by a factor of $N_1$, the number of
the nuclear spins in the polarized core, at a relatively small total
polarization. This protection effect may occur in quantum dots and solid-state spin systems defect centers, such as NV centers in diamonds, and could be harnessed to fabricate in a relatively simple way long-lived qubits and quantum memories.
\end{abstract}


\maketitle


{\it Introduction.}---Controlling and extending the coherence time of a qubit lie at the heart of spintronics, quantum computation, and quantum information processing. Decoherence occurs inevitably because of the interaction of the system with its
environment, which eventually makes the system behave classically. To counter
with the decoherence, many proposals have been developed, including dynamical
decoupling~\cite{Viola99}, decoherence-free subspace~\cite{Duan97, Lidar98}, and environmental state preparation~\cite{Klauser06, Stepanenko06, Takahashi08}.

Isolated electron spins in solids are promising information processors in
spintronics and quantum computing~\cite{Loss98, Zutic04}, due to their long  relaxation time in high magnetic fields as demonstrated in systems of quantum dots (QDs)~\cite{Fujisawa02}. 
In low magnetic fields, however, rapid relaxation may be caused by flips with
surrounding nuclear spins, typically in a timescale of 10 ns in a few mT
magnetic field and at $\sim$100 mK low temperature~\cite{Johnson05, Merkulov02,
Zhang06, Deng06}, which is not significantly longer than the nanosecond
operation clock in electrical control~\cite{Petta05, Koppens05}. Thus extension
of electron spin coherence time is highly desired~\cite{Greilich07}. A
particularly promising scheme is to prepare the nuclear spins in
low-fluctuation states~\cite{Klauser06, Stepanenko06, Ramon07, Ribeiro09, Reilly08, Xu09, Latta09, Vink09}, which also supplements to the dynamical
decoupling method by alleviating the requirements of pulse
control.

Nuclear spin state preparation has been investigated both theoretically and
experimentally. Previous investigations show that a nearly 100\% nuclear spin
polarization, which is difficult to realize in experiments, is required in
order to extend significantly the coherence time~\cite{Burkard99, Coish04, Deng06,
Takahashi08}, if a strong magnetic field is employed to uniformly polarize the nuclear spins. By contrast, the electron spin coherence time can be extended 10-100 times
by dynamically polarizing nuclear spins using repeated injection of polarized
electron spins at a rather low total nuclear polarization of $\sim 1\%$~\cite{Brown96, Bracker05, Ono02, Reilly08, Reilly08a, Petta08, Xu09}. But fully understanding the microscopic mechanisms of the dynamic nuclear polarization (DNP) of the coupled electron-nuclear spin systems remains a major challenge.

\begin{figure}
\includegraphics[width=3.25in]{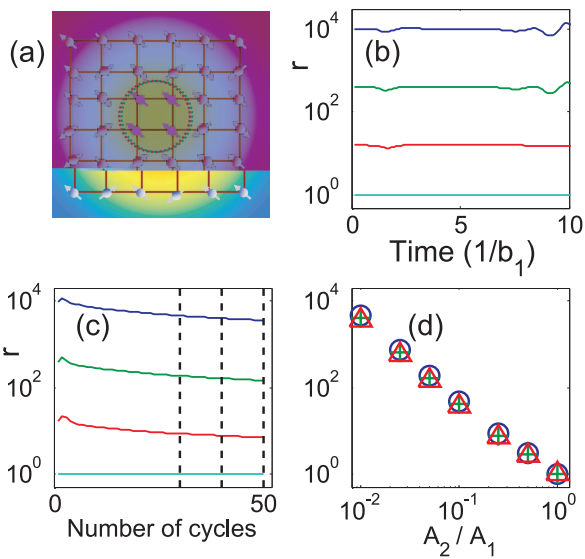}
\caption{\label{fig:dnp} (Color online) (a) Diagram of inhomogeneous hyperfine coupling in a quantum dot and the polarized nuclear spin core (spins inside the dashed circle). (b) Time dependence of the polarization ratio $r$ in a single DNP cycle for different $A_2 = 0.01, 0.05, 0.25, 1$ from top to bottom. Other parameters are $\omega_0 = 6$, $N_1=8$, $N_2=16$, $A_1 = 1$. The time unit is $1/b_1$ with $b_1=\sqrt{N_1 A_1^2}\;$. (c) Polarization ratio $r$ versus the number of DNP cycles. Other parameters are the same as in (b). Vertical dashed lines denote the cycle number $30,40,50$. (d) Polarization ratio $r$ versus hyperfine interaction strength ratio $A_2/A_1$ at DNP cycle $30$ (circles), $40$ (crosses), $50$ (triangles).}
\end{figure}

Aiming at understanding the evolution of the coupled electron-nuclear spin system, we present in this paper a microscopic picture of the formation of a
polarized core of nuclear spins during the DNP process by using a core-skirt two-region model [see Fig.~\ref{fig:dnp}(a)].
We focus on the protection of the electron spin coherence by the polarized
nuclear core in QDs, so as to provide a detailed and clear physical picture for
the DNP effect. Our results show that the electron spin relaxation time can be
extended substantially with a relatively small nuclear spin polarization. In a
QD with inhomogeneous electron density, the strongly coupled nuclear spins
(core spins) are easier to be polarized during DNP than the weakly coupled ones
(skirt spins). Once the $N_1$ core spins are polarized, they form a compound
with the electron spin, which is decohered as a whole by the skirt spins. When
the compound exchanges one quantum of moment with the skirt spins, the center
spin moment is changed approximately $1/(N_1+1)$ fraction of a quantum. So the
center spin relaxation is roughly $N_1+1$ times slower (see Fig.~\ref{fig:np}).


{\it Formation of polarized core during DNP.}---We consider a coupled electron and nuclear spin system with a Hamiltonian
(Gaudin spin-star model)
\begin{eqnarray}
{\cal H} &=& \omega_0 S_z + {\bf S} \cdot \sum_{k=1}^N A_k {\bf I}_k,
\label{eq:h}
\end{eqnarray}
where $\omega_0$ is the Zeeman energy of the central electron spin ${\bf S}$
($S=1/2$) and ${\bf I_k}$ ($I_k=1/2, k=1,2,\cdots, N$) is the $k$th nuclear
spin~\cite{Zhang06, Taylor07, Merkulov02}. We set $\hbar=1$ for simplicity.
Specifically, for an electron in a QD, $A_k$ is the Fermi contact hyperfine
coupling constant, which is proportional to $|\phi({\bf x}_k)|^2$, the electron
density at the $k$th nucleus. Due to the inhomogeneity of $|\phi|^2$, $A_k$ is
non-uniform in general. Such an inhomogeneity of $A_k$ is important to the formation of polarized nuclear spin core during DNP.

To illustrate the mechanism of formation of the polarized nuclear spin core, we
first consider a two-region model in which the nuclear spins consist of a core
and a skirt region. The hyperfine interaction strength $A_1$ of $N_1$ core
nuclear spins is stronger than the strength $A_2$ of $N_2$ skirt nuclear spins
[see Fig.~\ref{fig:dnp}(a)]. So the Hamiltonian in Eq.~(\ref{eq:h}) can be written
as
\begin{eqnarray}
{\cal H} &=& \omega_0 S_z + A_1{\bf S \cdot \bf I}_1 + A_2{\bf S \cdot \bf
I}_2. \label{eq:h2}
\end{eqnarray}
Here ${\bf I}_{1(2)}=\sum_{k=1}^{N_{1(2)}} {\bf I}_k$ is the total spin of the
nuclei in the core (skirt) region.
This Hamiltonian conserves $I_1$, $I_2$, and the $z$-component of the total
spin $S_z+I_{1z} + I_{2z}$, which enables an exact formulation of
the DNP process.

Below we show that in a finite external magnetic field the polarization of the
core spins acquired during the DNP process is much larger than that of the
skirt spins, provided that $A_1\gg A_2$. We assume that the initial electron
spin state is spin-up and the initial nuclear spin state is maximally mixed,
$\rho_0 = 2^{-(N_1+N_2)}|\uparrow\rangle \langle \uparrow| \otimes {\bf 1}$
with ${\bf 1}$ being a unit matrix of dimension $2^{N_1+N_2}$. Given that the
electron-spin-mediated indirect coupling between ${\bf I}_1$ and ${\bf I}_2$ is
negligible in a magnetic field $\omega_0 \gg \sqrt{N_{1,2}}A_{1,2}$, the
nuclear spin polarization saturates at a value $p_{1,2} = \langle I_{1z,2z}
\rangle / N_{1,2} \propto (A_{1,2}/\omega_0)^2$ at long times, according to the
perturbation theory. Thus the polarization ratio $r = p_1/p_2$ is proportional
to the square of the local hyperfine coupling strength, i.e., $r \propto
(A_1/A_2)^2$. As shown in Fig.~\ref{fig:dnp}, exact numerical calculations
according to Eq.~(\ref{eq:h2}) agree with the perturbation theory results. To
initiate a new DNP cycle, we reset the electron spin to the up state, i.e., set
the system to $|\uparrow\rangle\langle\uparrow|\otimes{\rm Tr}_S[\rho(T)]$
where the partial trace is over the electron and $T$ is the cycle period. After
many DNP cycles, numerical results in Fig.~\ref{fig:dnp} show that the relation
$r \propto (A_1/A_2)^2$ holds even better due to the fact that the total
nuclear spin polarization (thus the effective $\omega_0$) increases with the
number of DNP cycles. Of course, if the number of DNP cycles goes to infinity
(much greater than $N$),  the maximum polarization of the nuclear spins is in
the order of $1/\sqrt{N}$ for uniform hyperfine coupling ($A_k$ is constant)
and reaches the order of 1 for inhomogeneous $A_k$s. Notwithstanding this
limiting situation, the relation $p_k \propto A_k^2$ usually holds in realistic
experiments since the number of DNP cycles is $\lesssim N$~\cite{Petta08,
Reilly08}.


{\it Coherence protection effect of the polarized core.}---From the physical picture of the two-region model, we deduce that when the
electron-mediated indirect coupling between nuclear spins is negligible, the
relation $p_k\propto A_k^2$ should hold for a spin-star model with general
inhomogeneous coupling. To verify this, we perform numerical simulations with a
Gaussian distribution of the hyperfine coupling coefficients $A_k$. Such a
distribution can be realized, e.g., in a QD with a harmonic trap potential. The
results (not shown) confirm the conclusion.

To investigate the effect of DNP on the electron spin relaxation, we simulate
the dynamics of a coupled electron-nuclear spin system with Gaussian
distributed $A_k$ [see Fig.~\ref{fig:dnp}(a)]. To take into account
the effect of nuclear spin polarization after many DNP cycles, we assume that
in the initial state each nuclear spin has a polarization of $p_k(\beta) = {\rm
tanh}(\beta A_k^2)$ with an adjustable parameter $\beta$ related to the inverse
spin temperature~\cite{Zhang06}. We have neglected the possible phase
correlations between nuclear spins built during the DNP process, which should
decay much faster than the non-equilibrium spin polarization. For $\beta
A_k^2\ll 1$, $p_k \propto A_k^2$, which reproduces the polarization ratio due
to DNP with cycle number $\lesssim N$. As $\beta$ approaches infinity, $p_k$
saturates at 1, which accords with the limiting cases of an infinite number of
DNP cycles.

We consider both small- and large-bath cases, with the number of nuclei $N=20$
and $256$, respectively. For $N=20$, an exact evaluation is obtained using the
Chebyshev polynomial expansion of the evolution operator $U = \exp(-it{\cal
H})$~\cite{Dobrovitski03, Zhang07r}; For $N=256$, the method is based on
P-representation of the density matrix~\cite{Al-Hassanieh06, Zhang07r}. For
$N=20$, we also compare the P-representation method with the exact solution,
and the results [Fig.~\ref{fig:fid}(c)] show that the P-representation method
provides a good approximation.

\begin{widetext}
\begin{center}
\begin{figure}[tb]
\includegraphics[width=5in]{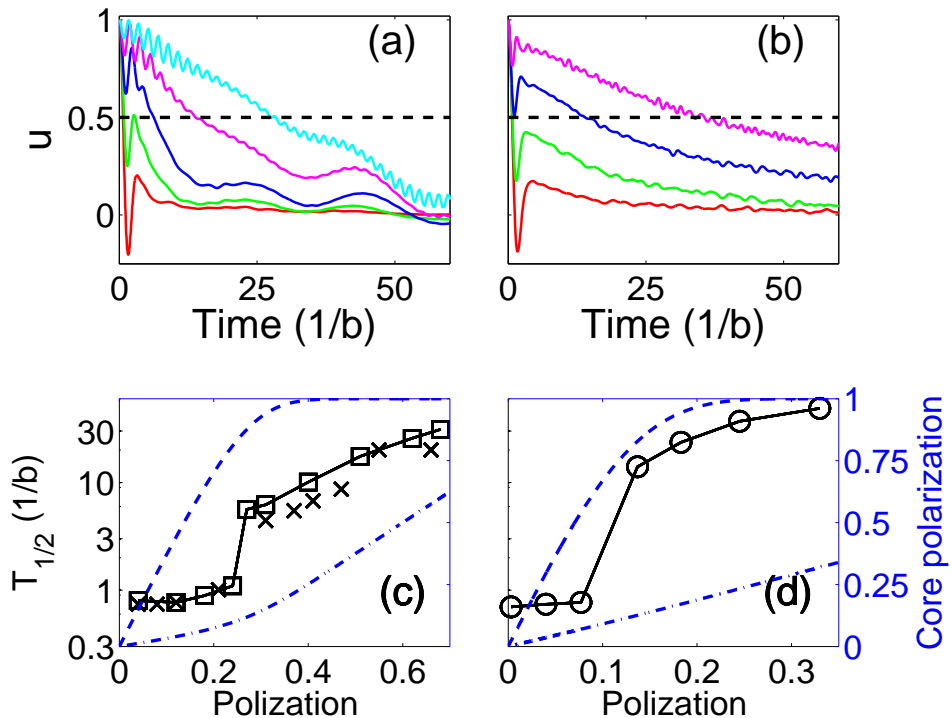}
\caption{\label{fig:fid} (Color online) Normalized decay of the electron spin polarization at various nuclear spin polarizations which increase from bottom to top for (a) $N=20$ and (b) $N=256$. The dashed horizontal lines mark the half decay point. The (last if multiple) cross point of the dashed horizontal line with each decay curve gives the half decay time $T_{1/2}$. (c) Dependence of $T_{1/2}$ (left $y$-axis) and the nuclear spin core polarization (right $y$-axis) on the nuclear spin polarization in the case of $N=20$. Coincidence of the formation of the polarized core and the jump of the relaxation time $T_{1/2}$ manifests the key role played by the core. Black solid line with squares --- the results of Chebyshev polynomial expansion method, black crosses --- the results of P-representation method, dashed lines --- core polarization, dash-dotted lines --- skirt polarization. (d) Same as (c) for $N=256$. Black solid line with circles --- the results of  P-representation method. The Gaussian width is $a$ for $N=20$ and $6.4a$ for $N=256$, respectively, with $a$ the lattice constant. The Gaussian center is shifted to (0.1, 0.27). }
\end{figure}
\end{center}
\end{widetext}

Figure~\ref{fig:fid} presents the relaxation of an electron spin from the
initial state $|\uparrow\rangle$. We set $\omega_0=0$ to rule out the external
magnetic field effect. The typical timescale is defined as
$b^{-1}=1/\sqrt{\sum_k A_k^2}$.  We show the results up to about $10^2\;b^{-1}$
since other mechanisms (such as dipolar nuclear spin interactions) need to be
included at longer times. The decay of the electron spin polarization is
normalized as
\begin{equation}
u \equiv \frac{\langle S_z(t)\rangle - \langle S_z(\infty)\rangle} {\langle
S_z(0)\rangle - \langle S_z(\infty)\rangle} \label{eq:norm}
\end{equation}
where $\langle S_z(t)\rangle = {\rm Tr}\{S_z\rho(t)\}$ with $\rho(t)$ the
density matrix of the whole system.

In Fig.~\ref{fig:fid}, significant extension of the relaxation time is observed
with increasing the nuclear spin polarization. For $N=20$, the decay time is
extended by about 40 times when the nuclear spin polarization is changed from 0
to 70\%. While for $N=256$, a similar extension requires less than 20\% nuclear
spin polarization. Such a large elongation of the decay time with a relatively
small polarization contrasts sharply to the case of thermally polarized nuclear
spins, where polarization as high as 90\% shows no significant extension of the
decay time~\cite{Zhang06, Deng08}. The contrast between the DNP and thermal
case indicates that the inhomogeneity in the nuclear spin polarization is of
key importance to the extension of the decay time. In Fig.~\ref{fig:fid}, we
also plot the polarization of 4 central nuclear spins in the core region. We see that
nearly full polarization of the core spins coincides with the abrupt increase
of the electron spin relaxation time. Actually, the factor of extension is
roughly the number of nuclei in the nearly fully polarized core region. This
phenomena suggest that the formation of a polarized core plays a critical role
in protecting the center spin coherence.

\begin{figure}
\includegraphics[width=3.25in]{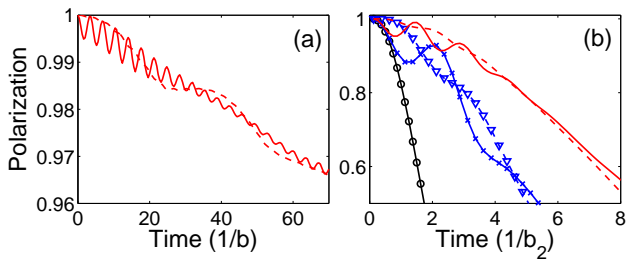}
\caption{\label{fig:np}(Color online) Nuclear spin core protection effect. (a) Solid
line denotes the electron spin polarization and dashed line
denotes core spin polarization for the $N=20$ case at 68\% total nuclear
polarization [the largest polarization data in Fig.~\ref{fig:fid}(c)]. (b) Two
region model results. Solid lines denote the electron spin polarization for
core spin number being $N_1=0$ (black line with circles), $2$ (blue line with crosses), and $4$ (red line). Dashed lines denote the core nuclear spin polarization for $N_1=2$ (blue dashed line with triangles) and $N_1=4$ (red dashed line). Other parameters are $A_1 = 1, A_2 = 0.1, N_2 = 40, \omega_0 = 0 $. The time unit is
$1/b_2$ with $b_2 = \sqrt{N_2 A_2^2}$.}
\end{figure}

To understand the ``protection'' effect, we resort to the previous two-region
model. If the core spins are fully polarized, they form together with the
center spin a compound of $N_1+1$ polarized spins at the state
$\left|\frac{N_1+1}2,\frac{N_1+1}2\right\rangle :=\left|S_z=\frac 1 2\right\rangle \otimes
\left|I_{1z}= \frac {N_1} 2\right\rangle.$ After a flip with a skirt spin, the state
of the core compound will be changed approximately to
$\left|\frac{N_1+1}2,\frac{N_1-1}2\right\rangle :=
\sqrt\frac{2N_1+1}{2(N_1+1)}\left|S_z=\frac 1 2\right\rangle \otimes \left|I_{1z}= \frac{N_1-1} 2\right\rangle
 + \frac 1 {\sqrt{2(N_1+1)}}\left|S_z=-\frac 1 2\right\rangle
\otimes \left|I_{1z}= \frac {N_1} 2\right\rangle.
$
Due to the hybridization of the
polarized core and the center spin, the electron spin is flipped with a
probability of about $1/(N_1+1)$. Thus the relaxation time is extended by a
factor of $(N_1+1)$.

To verify this picture, we examine the time-dependence of the nuclear core
polarizations during the electron spin relaxation process.
Figure~\ref{fig:np}(a), which is numerically calculated for $N=20$ with
Gaussian distributed coupling, shows that the core polarization decays in
accompany with the electron spin relaxation. The results for the 2-region model
shown in Fig.~\ref{fig:np}(b) are similar. For different number of core spins,
the electron spin decay time increases linearly with the number of core spins
$N_1$. All these features are consistent with the compound-spin picture.


{\it Discussion and conclusion.}---In a real QD system, the number of nuclear spins could reach as many as several  millions~\cite{Petta05}. According to the two-region model, 100 times extension of the coherence time requires a polarized core of about 100 nuclei, which corresponds to a total polarization of $\sim 0.01\%$. Considering the inhomogeneity of the hyperfine-interaction strength, we find it is very likely to extend $\sim 100$ times the electron-spin-coherence time with $\sim 1\%$ total polarization~\cite{Reilly08}. Full calculation of a double QD system will be presented in a future work.

In conclusion, we show that in an electron-nuclei spin system with inhomogeneous coupling, the DNP can lead to the formation of a highly polarized nuclear spin core, which in return suppresses the electron spin relaxation with a relatively low degree of total nuclear spin polarization. Such effect may be observed in quantum dots,  solid-state defect centers, and solid-state biomolecular NMR experiments~\cite{Ramanathan08}. In addition to the protection of the electron spin coherence, the polarized nuclear spin core is ready to be utilized to realize long-lived quantum memory based on imprinting and readout of electron spin state onto nuclear spins, which has a relaxation time on the scale of ten seconds~\cite{Reilly08a}, as proposed by Taylor and coworkers~\cite{Taylor03}. We anticipate that our results will also be of relevance in the full understanding of electron mediated nuclear spin diffusion during the DNP process of double QD systems~\cite{Reilly08a}.


{\it Acknowledgment.}---We thank Q. Y. Song and N. Zhao for discussions. This work was supported by the
China 973 Program grant No. 2009CB929300, NCET, and Hong Kong RGC Project CUHK 402208.



\end{document}